\documentclass[prl,twocolumn,showpacs,amsmath,amssymb]{revtex4-1}
\usepackage{graphicx,amsmath,bm, braket, txfonts, color}
\pdfoutput=1

\begin{document}

\title{Dynamic Magnetoelectric Effect in Ferromagnet$\mid$Superconductor Tunnel Junctions}
\author{Mircea Trif}
\author{Yaroslav Tserkovnyak}
\affiliation{Department of Physics and Astronomy, University of California, Los Angeles, California 90095, USA}
 
\date{\today}

\begin{abstract}
We study the magnetization dynamics in a ferromagnet$\mid$insulator$\mid$superconductor tunnel junction and the associated buildup of the electrical polarization. We show that for an open circuit, the induced voltage varies strongly and nonmonotonically with the precessional frequency, and can be enhanced significantly by the superconducting correlations. For frequencies much smaller or much larger than the superconducting gap, the voltage drops to zero, while when these two energy scales are comparable, the voltage is peaked at a value determined by the driving frequency.  We comment on the potential utilization of the effect for the low-temperature spatially-resolved spectroscopy of magnetic dynamics.
\end{abstract}

\pacs{76.50.+g, 74.25.F-, 72.15.Gd, 85.30.Mn}

\maketitle

%{\it Introduction|}

The field of spintronics has evolved  tremendously  over the last decades, leading to important conceptual and technological advances in spin-based memories, sensing, and logic \cite{MaekawaBook,*maekawaBOOK12}. The manipulation and detection of the spin degrees of freedom, such as collective magnetization, lies at the heart of spintronics, with magnetic field, static or time-dependent, providing a direct way to access it. It turns out, however, that the electric rather than magnetic control can be often preferred for spintronic manipulations \cite{ZuticRMP04}, as the former can exert larger torques, act faster, and can be applied or detected with a finer spatiotemporal resolution \cite{SlonczewskiJMMM96,*BergerPRB96,*kiselevNature2003,*RalphJMMM08,MaruyamaNatNano09,*ZhuPRL2012}. 

There are two main routes for the electrical control of magnetization dynamics:  One relies on spin-polarized electrical currents, which couple to the  magnetization via the spin-transfer torque \cite{SlonczewskiJMMM96}, while the other, which is also a more recent development, is based on controlling magnetic anisotropies by applying voltage pulses \cite{MaruyamaNatNano09}. Focussing on the former, the most basic  implementation for creating spin-polarized currents is to pass  unpolarized electrical current through a fixed ferromagnet.  An alternative way for generating spin currents relies on  the spin Hall effect, which leverages spin-orbit interaction in the material and does not require any  ferromagnetic polarizers.  In fact, this has been established as a primary tool both for manipulating and detecting the magnetization dynamics, transforming spin signals (spin currents) directly into electrical signals (Hall voltages) or vice versa \cite{KatoScience04,*ValenzuelaNat06,*SaitohAPL06,*KimuraPRL07,*AndoAPL11}.  The spin-Hall-induced voltage scales with the lateral dimension of the sample, making itself extremely useful for larger devices \cite{LiuScience12}. The main drawback, however, is that it loses its utility when it comes to detecting {\it local} magnetization dynamics.

In this Letter, we study voltage induced by magnetization dynamics in a circuit involving a driven metallic ferromagnet coupled to an $s$-wave superconductor via a weak tunnel barrier. It was previously shown that magnetization dynamics can induce voltage in tunnel junctions with normal metals or static reference ferromagnets \cite{MoriyamaPRL08,*XiaoPRB08,TserkovnyakPRB08}, by the process of an adiabatic charge pumping. The resulting voltage is, however, small, compared to the driving frequency, at typical microwave powers. We show here that singularity associated with the quasiparticle density of states can significantly enhance such dynamically induced voltages at driving frequencies corresponding to the superconducting gap. The underlying pumping behavior is thus necessarily nonadiabatic. 

\begin{figure}[t]
\begin{center}
\includegraphics[width=0.9\linewidth]{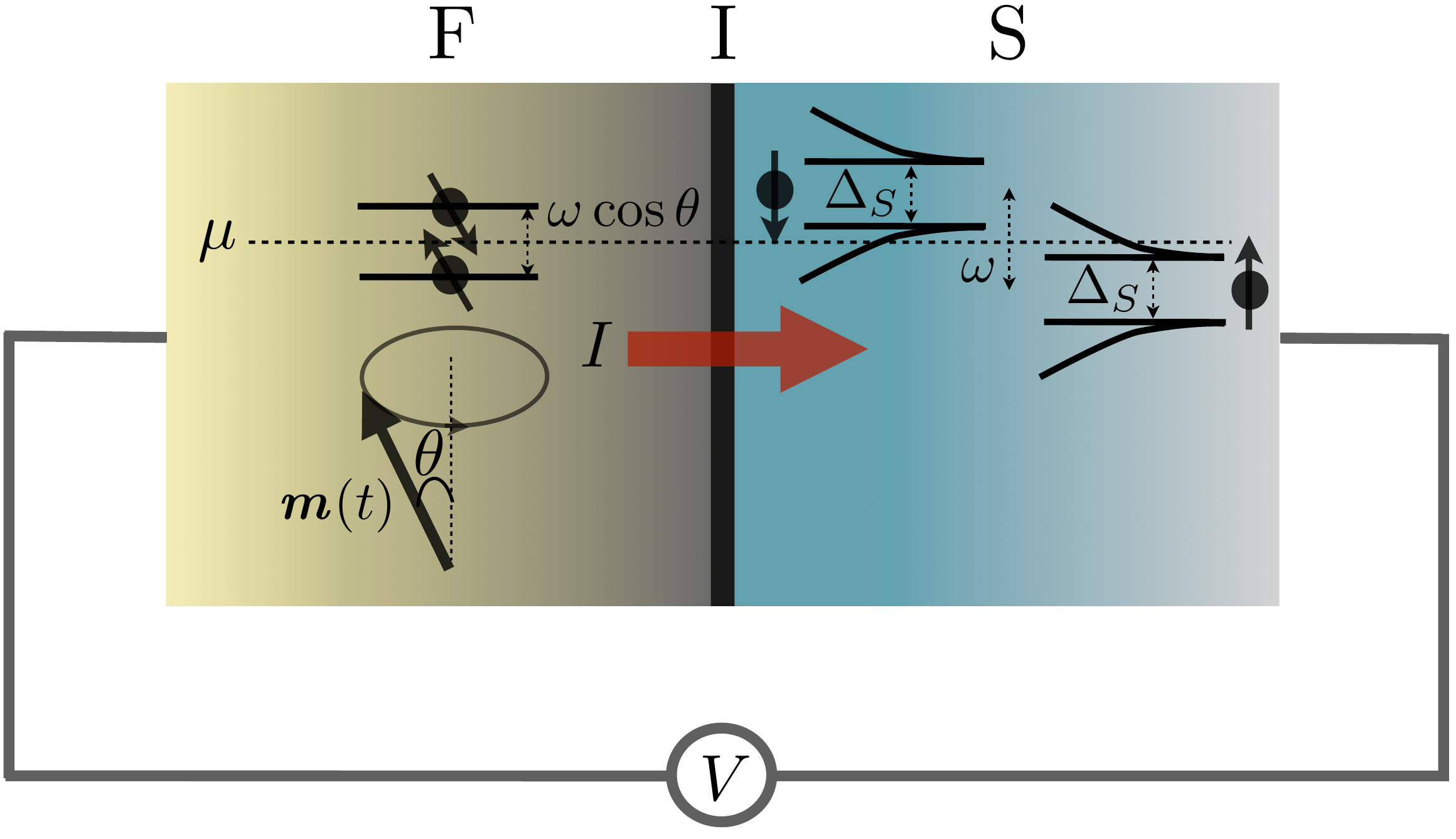}
\caption{A schematic of the system. A metallic  ferromagnet is driven to precess with frequency $\omega$ at angle $\theta$ with respect to the $z$ axis. The effective splitting in the rotating frame of reference is $\omega\cos\theta$ along the magnetization direction $\bm{m}$ within the ferromagnet. The superconductor is subjected to the effective spin splitting $\omega$ along the $z$ axis, which shifts the quasiparticle bands in the rotating frame. $\Delta_S$ is (twice) the $s$-wave superconducting gap and $\mu$ is the chemical potential in equilibrium in the laboratory frame. The magnetic precession pumps charge current $I$ through the tunnel barrier in the closed circuit, or a voltage $V$ is measured by a voltmeter in the open circuit. }
\label{sketch_system}
\end{center}
\end{figure} 

The Bogoliubov-de Gennes Hamiltonian for our hybrid ferromagnet/superconductor junction (see Fig.~\ref{sketch_system}) reads \cite{BrataasTserkovnyakPRLSCF04}
%{\it The Hamiltonian|} The Hamiltonian describing the \rm FIS structure reads:
\begin{align}
\hspace{-2mm}H_{\rm \rm FIS}(t)&=\left[\frac{p^2}{2m}+V(\bm{r})\right]\tau_z+\frac{\Delta_F(\bm{r})}{2}\bm{m}(t)\cdot\bm{\sigma}+\frac{\Delta_S(\bm{r})}{2}\tau_x\,,
\label{Ham}
\end{align}
in a certain basis, where $\bm{m}(t)$ is the magnetization direction in the ferromagnet, $\Delta_F(\bm{r})=\Delta_F\Theta(-x)$ is the magnetic (Stoner or \textit{s-d}) exchange field, $\Delta_S(\bm{r})=\Delta_S\Theta(x)$ is the (real-valued) superconducting pair potential, both written in terms of the Heaviside step function $\Theta(x)$, $\bm{\sigma}=(\sigma_x,\sigma_y,\sigma_z)$ and $\bm{\tau}=(\tau_x,\tau_y,\tau_z)$ are Pauli matrices operating in spin and particle-hole (Nambu) subspaces, respectively, and $V(\bm{r})$ is the total effective scalar potential acting on electrons.  We suppose the ferromagnetic and superconducting regions are separated by an insulating barrier, such that $V(\bm{r})$ is large near $x\approx0$, where $x$ stands for direction normal to the junction placed at $x=0$.

For a circular precession, we parametrize the magnetization direction as $\bm{m}(t)=[\sin{\theta}\cos{(\omega t)},\sin{\theta}\sin{(\omega t)},\cos{\theta}]$, with $\omega$ being the precession frequency and $\theta$ the precession angle. The exchange splitting $\Delta_F$ is assumed to be much larger than both the superconducting gap $\Delta_S$ and the precession frequency $\omega$: $\Delta_{F}\gg \Delta_{S},\omega$ (setting $\hbar=1$ throughout), which is typically the case. For a steady precession, it is convenient to switch to a rotating frame of reference, where the ferromagnet is static.  This is achieved by a time-dependent unitary transformation of the Hamiltonian: $H_{\rm \rm FIS}'\equiv U^{\dagger}(t)H_{\rm FIS}(t)U(t)-iU^{\dagger}(t)\partial_tU(t)=H_{\rm FIS}(0)-\omega\sigma_z/2$, with $U(t)=\exp{(-i\sigma_z\omega t/2)}$ \cite{TserkovnyakPRB08,BenderPRB10}. In the ferromagnetic bulk, the Hamiltonian in the rotating frame becomes $H_{F}'=H_{F}(0)-(\omega/2)\sigma_{\parallel}\cos{\theta}$, where $\sigma_{\parallel}\equiv\bm{m}(0)\cdot\bm{\sigma}$ is the spin projection on the magnetization direction $\bm{m}(0)$. We have disregarded the component perpendicular to $\bm{m}(0)$, which is effectively suppressed for $\omega/\Delta_F\ll1$. On the superconducting side, on the other hand, the spin splitting in the rotating frame is simply given by $\omega$, with the corresponding Hamiltonian $H_{\rm SC}'=H_{\rm SC}(0)-\omega\sigma_z/2$. The spectrum of the superconductor in the rotating frame is thus shifted by the effective magnetic (Larmor) field, $E_S^{(\sigma)}(\epsilon)=\pm\sqrt{\epsilon^2+(\Delta_S/2)^2}-\omega\sigma/2$, corresponding to the quasiparticle density of states
\begin{equation}
D_S^{(\sigma)}(E)=D_0\frac{|E+\omega\sigma/2|}{\sqrt{(E+\omega\sigma/2)^2-(\Delta_S/2)^2}}\,.
\label{DOS}
\end{equation}
Here, $D_0$ is the Fermi-level density of states per spin projection in the normal state, and $\sigma=\pm$ for spins up/down along the $z$ axis. $D^{(\sigma)}_S(E)$ diverges at energies $E\rightarrow(\pm\Delta_S-\omega\sigma)/2$. Since the spin-up and -down quasiparticle sub-bands in the superconductor are shifted by $\omega$, the gap closes in the rotating frame for $\omega>\Delta_S$. The spin-dependent spectrum on the ferromagnetic side is shifted too, but by $\omega\cos{\theta}$ instead of $\omega$. In the tunneling regime, we can assume that the ferromagnetic and superconducting bulks are in their separate equilibria in the laboratory frame. This relies on the fact that the magnetization-dynamics induced pumping is exponentially weak, such that, in particular, it has essentially no effect on the self-consistent pairing potential $\Delta_S$.

We now rewrite Hamiltonian \eqref{Ham} in the tunneling approximation, which is given in the second-quantized form by
\begin{equation}
H_T=\sum_{\bm{k},\bm{q},\sigma}t_{\bm{k},\bm{q},\sigma}c^{\dagger}_{F;\bm{k},\sigma}c_{S;\bm{q},\sigma}+{\rm H.c.}\,,
\label{TunHam}
\end{equation}
in the rotating frame. Here, $t_{\bm{k},\bm{q},\sigma}\equiv t$ is the tunneling matrix element, which is taken to be constant, for simplicity, and $c_{F(S);\bm{k},\sigma}$ are the electron annihilation operators in the ferromagnet (superconductor). $\bm{k}$ and $\bm{q}$  label orbital quantum numbers and $\sigma$ spin projection on the magnetic direction. We do not expect a general spin- and $\bm{k},\bm{q}$-dependent tunneling to affect qualitative features of our final results, apart from modifying parameters associated with the spin-dependent density of states in the ferromagnet. Hamiltonian (\ref{TunHam}) can be used to calculate the charge and spin currents flowing from one metal to the other across the tunnel barrier. 

We compute the out-of-equilibrium charge current using Fermi's golden rule for the transition probabilities involving all electron- and hole-like branches in the superconductor. In the tunneling regime, the superconductor can be effectively viewed as a simple  semiconductor \cite{BTKPRB82,TinkhamBook} (spin split in the rotating frame), with a singular quasiparticle density of states at its band edges, according to Eq.~(\ref{DOS}). The total current induced by the magnetic driving in the presence of a voltage $V$ reads:
\begin{align}
I=&2\pi e|t|^2\sum_{\sigma\sigma'}\int dE D_F^{(\sigma)}D_S^{(\sigma')}(E)\left|\langle\sigma|\sigma'\rangle\right|^2\nonumber\\
&\times\left[f_F^{(\sigma)}(E-eV)-f_S^{(\sigma')}(E)\right]\,,
\label{CurrentGeneral}
\end{align}
where $V$ is the electrochemical potential (applied and/or induced) of the ferromagnet relative to the superconducting condensate, $e<0$ is the electron's charge, $D_{F}^{(\sigma)}$ is the spin-$\sigma$ Fermi-level density of states in the ferromagnet, and $f_{F}^{(\sigma)}(E)=\{\exp{[(E+\omega\sigma\cos{(\theta)}/2)/k_BT]}+1\}^{-1}$, $f_{S}^{(\sigma)}(E)=\{\exp{[(E+\omega\sigma/2)/k_BT]}+1\}^{-1}$ are, respectively, the Fermi-Dirac distributions in the ferromagnet  and superconductor at temperature $T$. For the spin matrix elements, $\sigma$ labels spin in the ferromagnet along  $\bm{m}(0)$ and $\sigma'$  in the  superconductor along $\bm{z}$, so that  we have $|\langle\sigma|\sigma'\rangle|^2=\cos^2{(\theta/2)}\delta_{\sigma\sigma'}+\sin^2{(\theta/2)}\delta_{\bar{\sigma}\sigma'}$, where $\bar{\sigma}\equiv-\sigma$.  
We are now equipped to calculate the resultant current in the presence of the magnetization dynamics, at an arbitrary temperature. In contrast to band semiconductors, the gap $\Delta_S\to\Delta_S(T)$ itself depends on $T$, closing at $T_c\approx1.76\Delta_S(0)$ within the $s$-wave BCS model \cite{TinkhamBook}, while the Fermi level is pinned, in the laboratory frame, midgap by the superconducting condensate.

We start by computing the charge current as a function of $\omega$ and  $V$ at $T=0$. This regime allows for an analytical evaluation of the current, as well as captures qualitative features that extrapolate to finite $T$. Writing $I=\sum_{\sigma\sigma'}I_{\sigma\sigma'}$, we arrive at:
\begin{align}
I_{\sigma\sigma'}=&I^{(\sigma)}_0(1+\sigma\sigma'\cos{\theta})\sqrt{\left[\omega(\sigma -\sigma'\cos{\theta})/2-V\right]^2-1}\nonumber\\
&\times\left\{\sum_{s=\pm1}\Theta[\omega(\sigma-\sigma' \cos{\theta})/2-V+s]-1\right\}\,,
%I_{\sigma\sigma'}=&I^{(\sigma)}_0(1+\sigma\sigma'\cos{\theta})\bigg\{\sqrt{\left[V+\omega(\sigma'+\sigma\cos{\theta})/2\right]^2-1}\left(\Theta\left[V+\omega(\sigma'+\sigma\cos{\theta})/2-1\right]-\Theta\left[-V-\omega(\sigma'+\sigma\cos{\theta})/2-1\right]\right)\nonumber\\
%&-\sigma'\sqrt{\omega^2-1}\Theta(\omega-1)\bigg\}\,,
\label{TZeroCurrent}
\end{align} 
where $I^{(\sigma)}_0=\pi e\Delta_S(0)|t|^2 D_F^{(\sigma)} D_0$. Here, we are measuring all the energies in units of the zero-temperature gap, $\Delta_S(0)/2$. In the following, we focus on the open-circuit case (see Fig. 1), so that the current flowing through the heterostructure is zero ($I=0$), while the voltage $V$ induced by the magnetization dynamics is measured by a voltmeter. We are specifically interested in regimes where this voltage can be significantly enhanced by the superconductor. In the tunneling regime, the Andreev processes are strongly suppressed, so that transport is governed by the excited quasiparticles \cite{NoteRichardPRL12}. This means that for $\omega<1$ (in units of $\Delta_S/2$) and $T=0$, the quasiparticles are gapped out and $V\equiv0$, as long as the microwave power is low enough so that multimagnon processes do not contribute \cite{note1}. 

\begin{figure}[t]
\begin{center}
\includegraphics[width=0.9\linewidth]{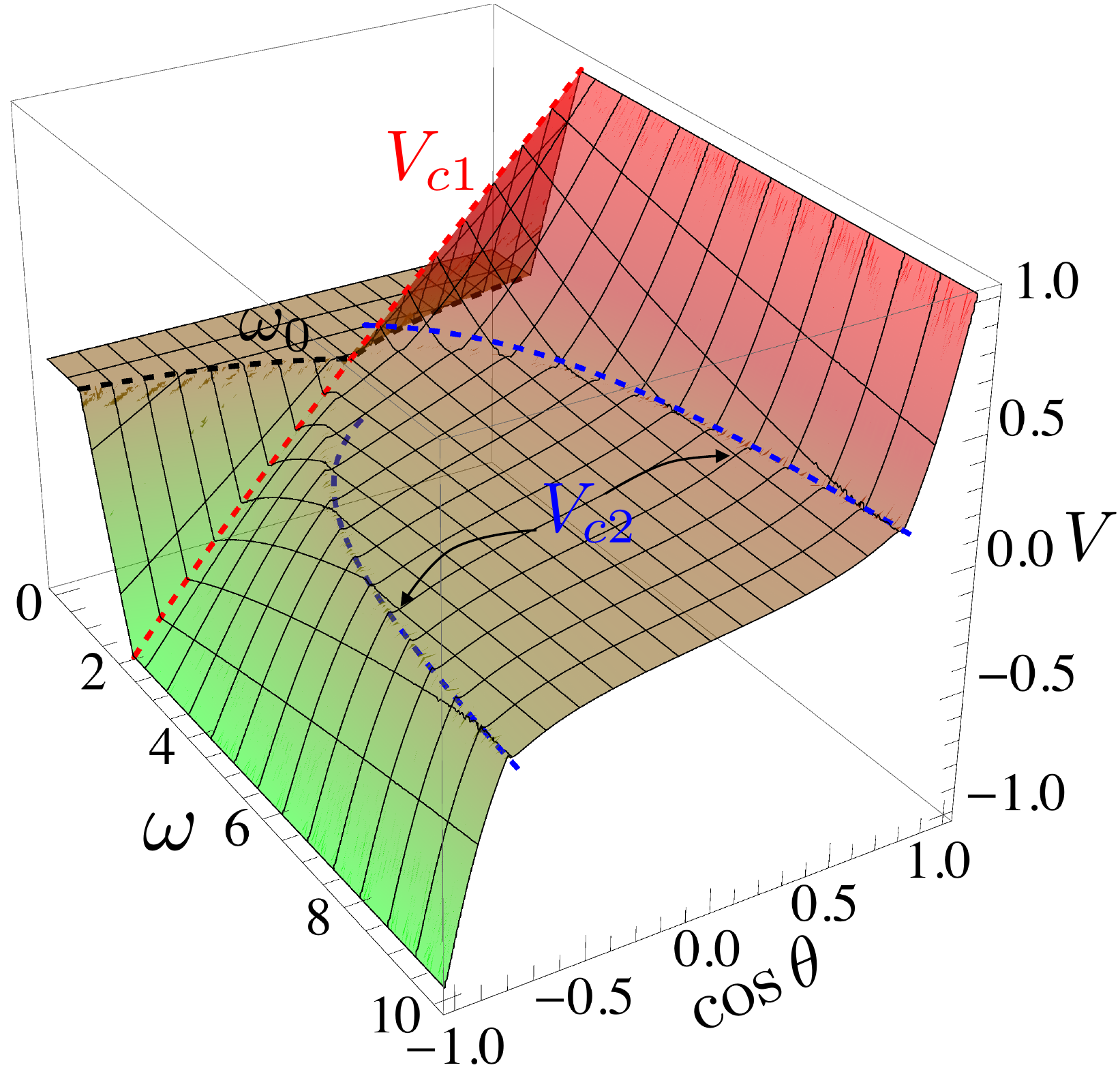}%VdepTZero.pdf}
\caption{Dependence of the induced voltage $V$ on frequency $\omega$  and $\cos{\theta}$ at zero temperature. The thick black curve shows the activation frequency $\omega_0=2/(1+\cos{\theta})$, while the red and blue curves correspond respectively to the voltages $V_{c1}$  [Eq. (\ref{Vcr1})]  and $V_{c2}$  [Eq. (\ref{Vcr2})] that signal the activation of new transport channels in the current $I$. In these plots, we set $P=2/3$ and expressed $V$ and $\omega$ in units of $\Delta_S/2$.}
\label{VdepTzero}
\end{center}
\end{figure}

In Fig.~\ref{VdepTzero}, we plot the open-circuit voltage $V$ at $T=0$, as a function of frequency $\omega$  and precession angle $\theta$. We find three different regimes in the dependence of $V$ on $\omega$, corresponding to the activation of new spin channels in the  tunneling current \eqref{TZeroCurrent} with increasing $\omega$. The voltage depends on $\omega$ nonmonotonically, reaching a maximum $V_{\rm c1}$ which, as shown later,  can be as large as $1$. More specifically,  we find that below a minimum frequency $\omega_{\rm 0}=2/(1+\cos{\theta})$ the voltage is always zero (assuming only positive frequencies, i.e., $\omega>0$), while for $\omega>\omega_{\rm 0}$ there is a finite voltage drop in the system. The  total current in this case is given initially by $I_{+-}+I_{-+}$ for $0<\theta<\pi/2$ and $I_{++}+I_{--}$ for $\pi/2<\theta<\pi$. At certain higher frequencies (whose specific values are discussed below), the terms $I_{++}$ and $I_{--}$ get activated for $0<\theta<\pi/2$ and $I_{+-}$ and  $I_{-+}$ for $\pi/2<\theta<\pi$. Successive activation of these different tunneling channels define voltage landscapes  as  shown in Fig.~\ref{VdepTzero}. Note that the symmetry of our system dictates that $V(-\omega)=-V(\omega)$ as well as $V(\pi-\theta)=-V(\theta)$, allowing us to henceforth restrict our discussion to $\theta\in(0,\pi/2]$ and $\omega>0$.  While it is possible to extract analytical expressions for the voltage (and the activation frequencies) as a function of frequency at $T=0$, these are too long and unilluminating. Instead, we will derive approximate expressions for the induced voltage in different frequency ranges. 

At frequencies $\omega$ above $\omega_{0}$ but still sufficiently low such that $|\omega\cos^2{(\theta/2)}\pm V|>1$ and  $|\omega \sin^2{(\theta/2)}\pm V|<1$ [where $V\equiv V(\omega,\theta,P)$ needs to be solved self-consistently for, in the open circuit], the voltage $V$ increases monotonically with $\omega$, as shown in Fig.~\ref{VdepTzero}, given approximatively [as an expansion in $(1-P)$] by the following expression:
\begin{align}
V\approx\left(\omega\cos^2{\frac{\theta}{2}}-1\right)\left[1-2\omega\left(\frac{1-P}{1+P}\right)^2\cos^2{\frac{\theta}{2}}\right]\,.
\label{Vreg1}
\end{align}
Here, we defined the polarization $P=(D_{F}^{\downarrow}-D_F^{\uparrow})/(D_{F}^{\downarrow}+D_F^{\uparrow})$ and assumed  $0<P\simeq1$ (large polarization). We see that in this limit  the voltage increases roughly  linearly with $\omega$, consistent with the exact result shown in  Fig.~\ref{VdepTzero}, until the frequency reaches a critical value $\omega_{\rm c1}=2[(1+P)^2+(1-P)^2\cos{\theta}]/[4P+(1-P)^2\sin^2{\theta}]$  corresponding to the condition $|\omega\sin^2{(\theta/2)}+V|=1$. At this frequency, the  term  $I_{++}$ starts contributing to the total current $I$, and the voltage starts decreasing monotonically with increasing $\omega$, as depicted in  Fig.~\ref{VdepTzero}. The critical voltage in the circuit at  $\omega=\omega_{\rm c1}$  reaches a value  $V_{\rm c1}$ given by
\begin{equation}
V_{\rm c1}=\frac{4P\cos{\theta}}{4P+(1-P)^2\sin^2\theta},
\label{Vcr1}
%\frac{(1-p^2)\cos{\theta}}{1-p^2\cos^2{\theta}}
\end{equation}
which approaches unity as $\theta\to0$. This is the maximum voltage achievable in the circuit, which, in physical units, is bounded by the superconducting gap $\Delta_S$/2. Note  that  $V$ generally increases with decreasing angle $\theta$, as shown in Fig.~\ref{VdepTzero}, being in stark contrast to the typical ${\rm F | I |F}$ magnetic junction with one ferromagnet being free and one pinned, where $V\propto\sin^2\theta$, i.e., vanishing with microwave power, as the precession angle $\theta\rightarrow0$. Moreover,  in the present setup, the voltage does not depend on the orientation of the precession axis,  as opposed again to the ${\rm F| I |F}$, where  the induced voltage is sensitive to the relative orientation of the axis of precession with respect to the pinned reference ferromagnet \cite{TserkovnyakPRB08,XiaoPRB08}.

For $\omega>\omega_{\rm c 1}$, we find that $V$ decreases with increasing $\omega$, and it is given approximatively  by the following expression:
\begin{align}
V&\approx\sqrt{1+\left(\frac{2P}{1+P}\omega\sin^2{\frac{\theta}{2}}\right)^2}-\omega\sin^2{\frac{\theta}{2}},
\label{Vreg2}
\end{align}
where we neglected corrections of the order of $1/\omega\cos^2{(\theta/2)}\ll1$. Moreover, $\omega\sin^2{(\theta/2)}\sim1$,  since, in this regime, $|\omega\sin^2{(\theta/2)}+V|>1$ and $|\omega\sin^2{(\theta/2)}-V|<1$, while $0<V<1$ for all frequencies $\omega$.

As the frequency increases further beyond $\omega_{\rm c1}$, the voltage continues to decrease according to Eq.~\eqref{Vreg2} until the term $ I_{--}$ in the total current $I$ is activated, which happens when $|\omega\sin^2{(\theta/2)}-V|=1$. This determines the second transition frequency, $\omega_{\rm c2}\approx(1+P)^2/(1+2P)\sin^2(\theta/2)$, at which the voltage is given by
\begin{align}
V_{\rm c2}&\approx \frac{P^2}{2P+1},
\label{Vcr2}
\end{align}
neglecting corrections of order $1/\omega\cos^2{(\theta/2)}$. According to Eq.~\eqref{Vcr2}, the maximum voltage at this transition point is $V_{\rm c2}\approx1/3$, corresponding to $P=1$.  

\begin{figure}[t]
\begin{center}
\includegraphics[width=0.95\linewidth]{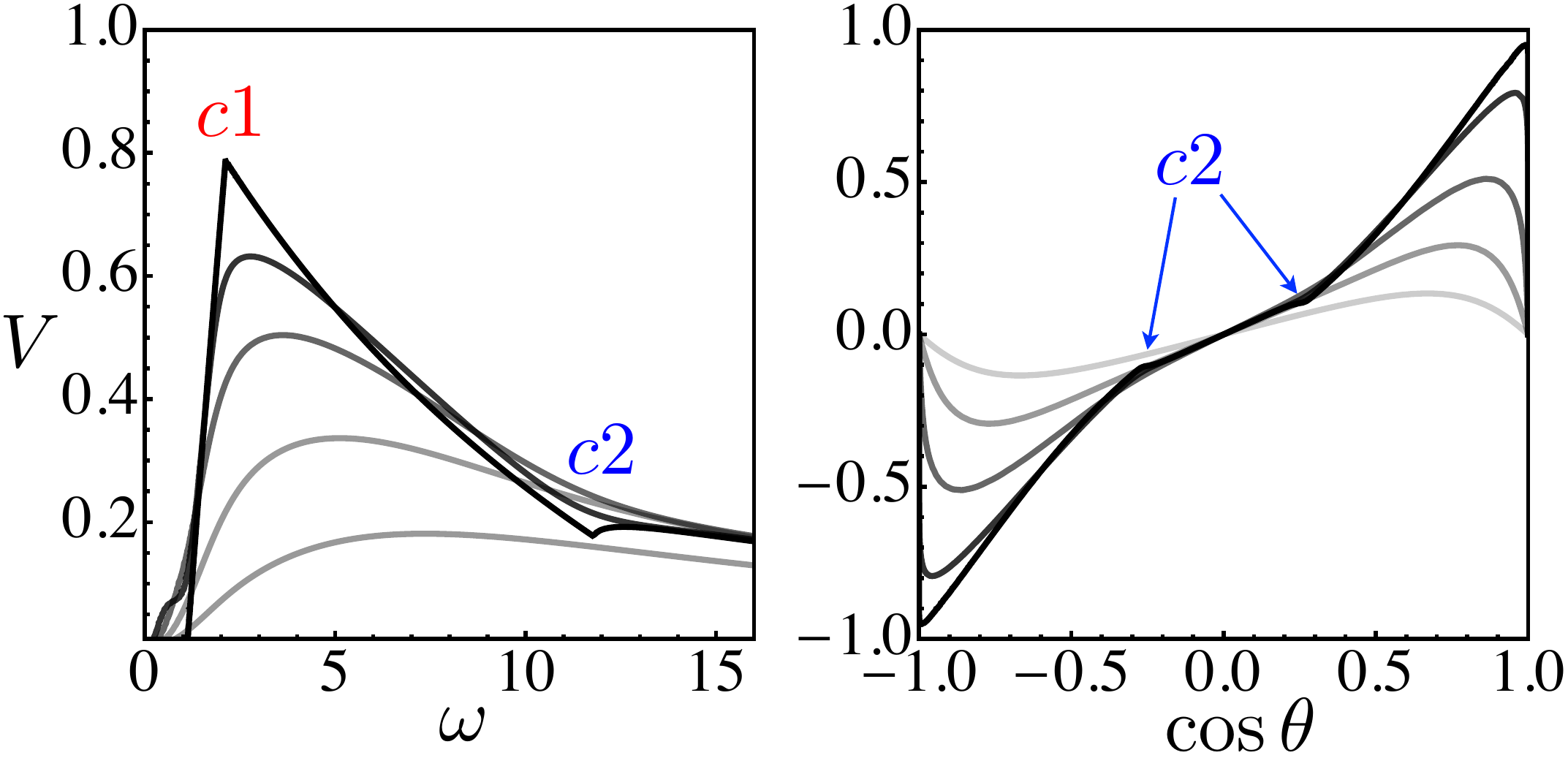}
\caption{Dependence of the induced voltage $V$ on  $\omega$ (left panel), and $\cos{\theta}$  (right panel) for different temperatures. The increasingly lighter gray  curves  correspond to $T/T_c=0.1, 0.3, 0.5, 0.7$, and $0.9$ with  $\cos{\theta}=0.8$ (left)  and  $\omega=3$ (right).  Here  $c1$ and $c2$  label the critical points at which new charge transport channels are activated [see Eq.~(\ref{Vcr1}) and Eq. (\ref{Vcr2})] at $T=0$.  In these plots, we set  $P=2/3$, and expressed $V$ and $\omega$ in units of $\Delta_S(0)/2$.}
\label{VdepFiniteT}
\end{center}
\end{figure} 

Finally, for $\omega>\omega_{\rm c2}$, all terms in Eq.~\eqref{TZeroCurrent} contribute to the charge current $I$, and the voltage tends to zero as $\omega\to\infty$. Specifically, for $\omega\gg\omega_{\rm c2}$, we find the following approximation for the voltage:
\begin{equation}
V\approx \frac{P}{\omega \sin^2{(\theta/2)}}\,.
\label{Vreg3}
\end{equation}
The vanishingly small voltage at large $\omega$ reproduces the nil result of ferromagnet${\mid}$normal-metal junctions \cite{TserkovnyakPRB08}, since the  superconducting correlations become unimportant at frequencies $\omega\gg\Delta_S$. 
%In the right panel of Fig.~\ref{VdepTzero}, we plot the angle dependence of the induced voltage for different values of $\omega$. The voltage oscillates as a function of $\theta$, such that $V(\pi-\theta)=-V(\theta)$, and  $V(0)=V(\pi/2)=V(\pi)\equiv0$. Also, the two maxima for $V$ are pushed from $\theta\approx\pi/2$ at $\omega\approx1$, to $\theta=0$ and $\pi$ with increasing $\omega$, with the peak voltage achieving a maximum value at an intermediate frequency of order 1. 

\begin{figure}[t]
\begin{center}
\includegraphics[width=0.95 \linewidth]{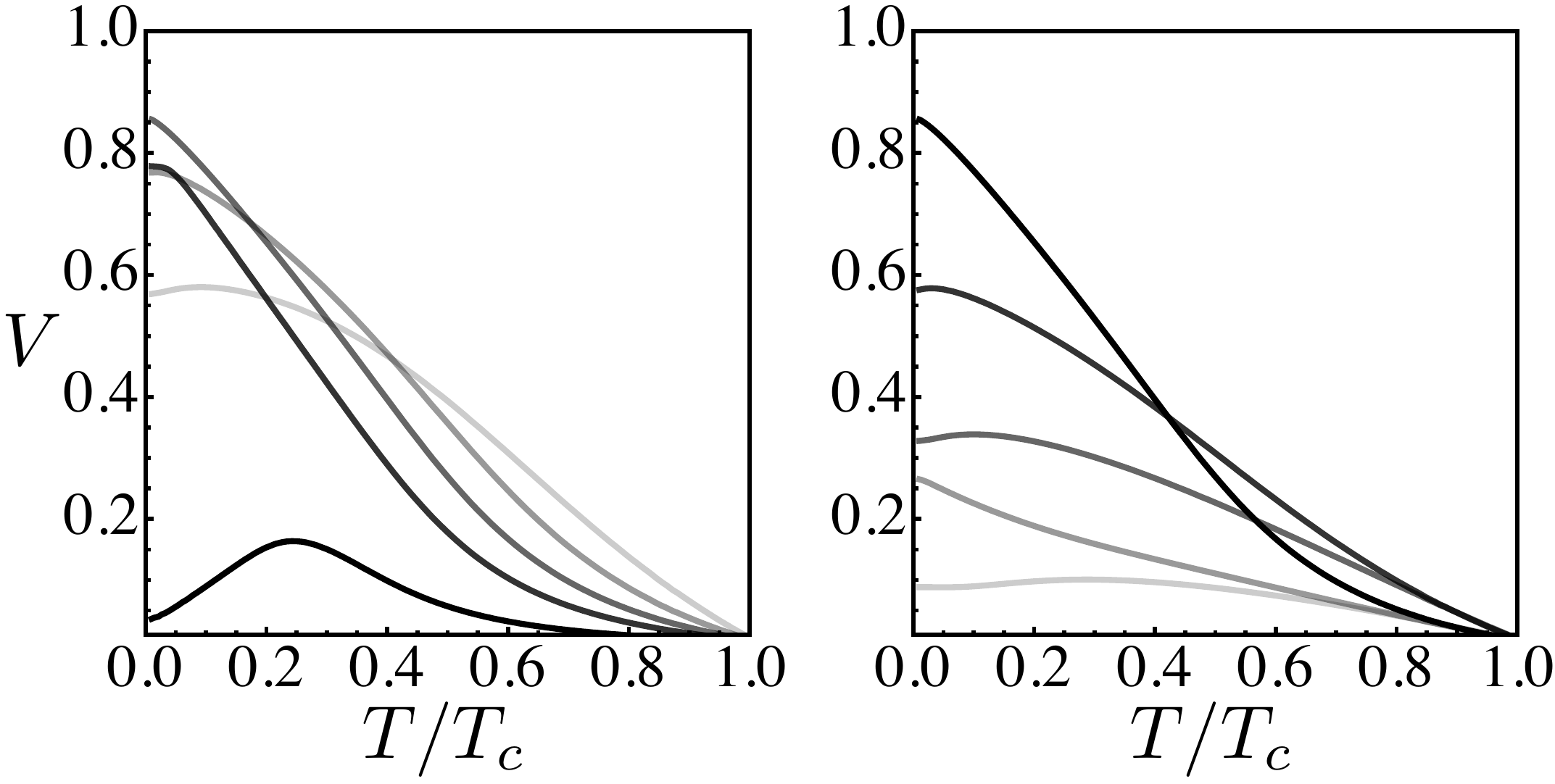}
\caption{Dependence of the induced voltage $V$ on temperature for different values of $\omega$ (left panel) and angle $\theta$ (right panel). Left: The increasingly lighter gray  curves  correspond to $\omega=1.1, 2, 3, 5$, and $10$,  with $\cos{\theta}=0.9$. Right: The increasingly lighter gray  curves  correspond to $\cos{\theta}=0.9, 0.7, 0.5, 0.3$, and $0.2$, with $\omega=3$. In these plots, we set $P=2/3$ and expressed $V$ and $\omega$ in units of $\Delta_S(0)/2$.}
\label{VdepofT}
\end{center}
\end{figure} 

Note that at each transition point from one regime to another, the slope of $V$ with respect to $\omega$ is discontinuous, as opposed to the case of the normal magnetic junctions, where it is constant for typical microwave frequencies $\omega\ll\Delta_F$ \cite{TserkovnyakPRB08}. The resulting voltages are limited by the driving frequency, achieving values $\sim\omega$ when $\omega\sim1$ (or $\omega\sim\Delta_S/2$, in physical units) which  are well within the experimental reach for typical superconducting gaps on the order of a few Kelvins, corresponding to a fraction of an meV. This is significantly larger than $\sim1~\mu$eV voltages induced by magnetic dynamics in normal metals \cite{MoriyamaPRL08}. 

At finite temperatures, we expect rounding off of the sharp features present at the transitions between different aforementioned regions as well as a reduction of the induced voltage, due to the diminished superconductivity. In order to calculate explicitly the pumped current at $T\neq0$ using Eq. (\ref{CurrentGeneral}), we need to account not only for the thermally broadened Fermi distribution but also for the $T$ dependence of the gap $\Delta_S(T)$, which closes at $T_c$. We write $\Delta_S(T)=\Delta_S(0)F(T)$, where $F(T)$ is a dimensionless function, which can be found numerically from the self-consistent gap equation in the BCS theory \cite{TinkhamBook}.  In Fig.~\ref{VdepFiniteT}, we plot the corresponding dependence of $V$ on $\omega$ (left) and $\cos{\theta}$ (right) at different temperatures, with the black curves showing the $T=0$ result. We see that the signal becomes visibly reduced as the temperature increases from zero and, moreover, it singularly changes the $\cos{\theta}\rightarrow1$ behavior such that $V\rightarrow0$ for any $T\neq0$ instead of a finite $V\to1$ at precisely $T=0$.   

%At $T=T_c$, the superconductor turns into a normal metal, and the induced voltage should vanish within our model, which is precisely the case as can be seen in the right panel of  Fig.~\ref{VdepFiniteT}.

In the left panel of Fig.~\ref{VdepofT}, we plot the induced voltage as a function of $T$ for different frequencies $\omega$ (left) and $\cos{\theta}$ (right).  As expected, all sharp features are smoothed out by finite $T$, with the signal eventually  vanishing as $T\rightarrow T_c$. At subcritical temperatures, the voltage shows a nonmonotonic behavior as a function of $T$, for a wide range of frequencies and angles, surprisingly reaching values in excess of the $T=0$ result. This is attributed to thermal activation of otherwise closed pumping channels. Had the gap been the same at all temperatures, the voltage would have exhibited an even more dramatic increase as the temperature is raised from $T=0$, which is suppressed by the thermal reduction of the gap.
 
In conclusion, we analyzed the dynamics  of a ferromagnet tunnel-coupled to a conventional superconductor. We find a large (compared to the normal junction) nonadiabatic electrical polarization induced in an open circuit, when $\omega\sim\Delta_S$. We speculate that it could be useful, in practice, as a local probe for magnetic dynamics. This voltage can be easily understood to stem from the fact that the superconductor effectively behaves as a static reference ferromagnet in the rotating frame, and we predict the maximum induced voltage of order of $\Delta_S$, for small  precession angles and temperatures $T<T_c$.

This work was supported in part by FAME (an SRC STARnet center sponsored by MARCO and DARPA), the NSF under Grant No. DMR-0840965, Grant No. 228481 from the Simons Foundation.

%\bibliographystyle{apsrev4-1}
%\bibliography{all}  

\begin{thebibliography}{22}
\bibitem{NoteRichardPRL12}
The charge transport  induced by the magnetization dynamics in the sub-gap regime, mediated by Andreev processes, was analyzed in Ref.~\onlinecite{RichardPRL12} in a setup where the ferromagnet is coupled to the superconductor by a metallic quantum dot.%
\bibitem{note1}
Although absent from our model with circular precession, we expect the multimagnon processes to become relevant for general elliptical precession at sufficiently high microwave powers.
\end{thebibliography}
%\end{document}

%merlin.mbs 2010-03-15 4.21a (PWD, AO, DPC)

\end{document}